\documentclass[aps,prc,superscriptaddress,nofootinbib,groupedaddress,12pt]{revtex4}
\usepackage{dcolumn}   
\usepackage{bm}        
\usepackage{multirow}

\usepackage{amsmath}
\usepackage{amsfonts}
\usepackage{amssymb}
\usepackage{makeidx}
\usepackage{graphicx}
\usepackage{anysize}
\usepackage{hyperref}

\usepackage{slashed}

\newenvironment{mymathbox}
{\par\smallskip\centering\begin{lrbox}{0}%
\begin{minipage}[c]{0.8\textwidth}}
{\end{minipage}\end{lrbox}%
\framebox[0.9\textwidth]{\usebox{0}}%
\par\medskip
\ignorespacesafterend}
\newcommand{\bb}{\begin{mymathbox}}
\newcommand{\eb}{\end{mymathbox}}

\newcommand{\be}{\begin{equation}}
\newcommand{\ee}{\end{equation}}
\newcommand{\ba}{\begin{eqnarray}}
\newcommand{\ea}{\end{eqnarray}}
\newcommand{\npsi}{{\bf \npsi}}

\newcommand{\bma}{\begin{pmatrix}}
\newcommand{\ema}{\end{pmatrix}}

\usepackage[dvips]{color}

\begin{document}

\title{Modeling neutrino-nucleus interaction at intermediate energies}

\author{R. Gonz\'alez-Jim\'enez}
\author{N. Jachowicz}
\author{A. Nikolakopoulos}
\author{J. Nys}
\author{T. Van Cuyck} 
\author{N. Van Dessel}
\affiliation{Department of Physics and Astronomy, Ghent University, Belgium}
\author{K. Niewczas}
\affiliation{Department of Physics and Astronomy, Ghent University, Belgium}
\affiliation{Institute of Theoretical  Physics, University of Wroc{\l}aw, Wroc{\l}aw, Poland}
       
\author{V. Pandey}
\affiliation{Center for Neutrino Physics, Virginia Tech, Blacksburg, Virginia, USA}
       
\begin{abstract}
  We present the current status of the research activities of the Ghent group on neutrino-nucleus interactions. 
These consist in the modeling of some of the relevant neutrino-nucleus reaction channels at intermediate energies: low-energy nuclear excitations, quasielastic scattering, two-nucleon knockout processes and single-pion production. 
The low-energy nuclear excitations and the quasielastic peak are described using a Hartree-Fock-CRPA (continuum random phase approximation) model that takes into account nuclear long-range correlations as well as the distortion of the outgoing nucleon wave function. 
We include two-body current mechanisms through short-range correlations and meson-exchange currents. Their influence on one- and two-nucleon knockout responses is computed. Bound and outgoing nucleons are treated within the same mean-field framework. 
Finally, for modeling of the neutrino-induced single-pion production, we use a low-energy model that contains resonances and the background contributions required by chiral symmetry. This low-energy model is combined with a Regge approach into a Hybrid model, which allows us to make predictions beyond the resonance region.\\

\small{The 19th International Workshop on Neutrinos from Accelerators-NUFACT2017,
25-30 September, 2017. Uppsala University, Uppsala, Sweden.}

\end{abstract}

\maketitle

\section{Introduction}

Neutrinos interact only weakly. Therefore, one needs as much matter as possible to detect them with the desired statistics. 
The use of complex targets made of medium-size nuclei, such as mineral oils (CH$_X$), water, or liquid argon, allows for the accumulation of tons of detector material, what significantly increases the statistics in neutrino detectors. As a consequence, past, current and next generations of neutrino experiments (MiniBooNE, MINERvA, T2K, MicroBooNE, DUNE, NOvA)~\cite{Alvarez-Ruso17} use `complex' nuclei as target material. This is what brings nuclear physics to the stage of neutrino-oscillation physics.

\begin{figure}[htbp]
  \centering  
      \includegraphics[width=0.65\textwidth,angle=0]{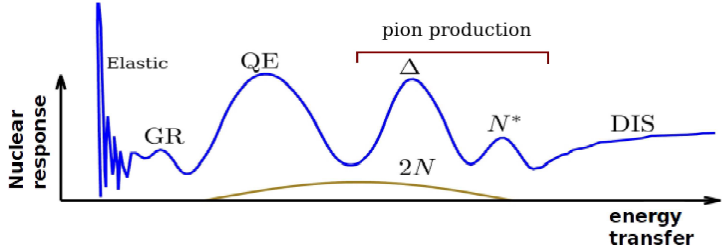}
  \caption{Electroweak nuclear response as a function of the energy transfer. The dominant channels are collective nuclear excitations, QE peak, pion production, DIS and a background from multinucleon contributions (dominated by two-nucleon knockout reactions, 2N). Figure adapted from~\cite{VanCuyck-PhD}.}
  \label{fig:spectrum}
\end{figure}

Systematic errors are a pivotal problem in the aforementioned neutrino-oscillation experiments. One of the most important sources of uncertainties is our poor knowledge of the neutrino-nucleus interaction. Currently, the neutrino-nucleus scattering cross sections, in the 1-5 GeV energy region (intermediate energies), are known with a precision not exceeding 20\%~\cite{Alvarez-Ruso17}. 
Another major problem is that the energy of the incident neutrino is unknown. This implies that any theoretical approach that aims at describing the current and forthcoming neutrino scattering data, has to contain all the essential ingredients of the cross section. 
At intermediate energies, the dominant reaction channels are (see Fig.~\ref{fig:spectrum}): low-energy nuclear excitations, giant resonances (GR), quasielastic (QE) scattering, multinucleon contributions, pion production, and deep-inelastic scattering (DIS).
The probability that one or the other reaction mechanism will take place depends on the energy transferred by the neutrino to the nucleus. This is sketched in Fig.~\ref{fig:spectrum}. 

\section{Models and Results}

In recent years, the research activities of the Ghent group have focused on providing a description of some of the neutrino-nucleus reaction mechanisms that are important at intermediate energies. In particular, we have focused on the modeling of the low-lying nuclear excitations, quasielastic scattering, two-body current contributions, and single-pion production. In what follows, we present an overview of our models and results.

\subsection{Giant Resonance region and Quasielastic peak}\label{GR-QE}

The nuclear ground state is described within a Hartree-Fock-CRPA approach, i.e., the wave functions of the bound nucleons are obtained by solving the Schr\"odinger equation with a self-consistent mean-field potential generated by an effective Skyrme nucleon-nucleon interaction. Long-range correlations, that account for collective nuclear effects in the giant resonance region, are introduced by a continuum random phase approximation (CRPA) approach, where the same Skyrme parameterization is used as interaction. 
The outgoing nucleon is under the influence of the nuclear potential, hence, elastic final-state interactions (FSI) are included.   
Other improvements to the model, such as relativistic corrections and a dipole form factor controlling the RPA strength at large $Q^2$, have been implemented and are discussed in~\cite{Pandey14,Pandey15,Martini16}.

\begin{figure}[htbp]
  \centering  
       \includegraphics[width=0.45\textwidth,angle=0]{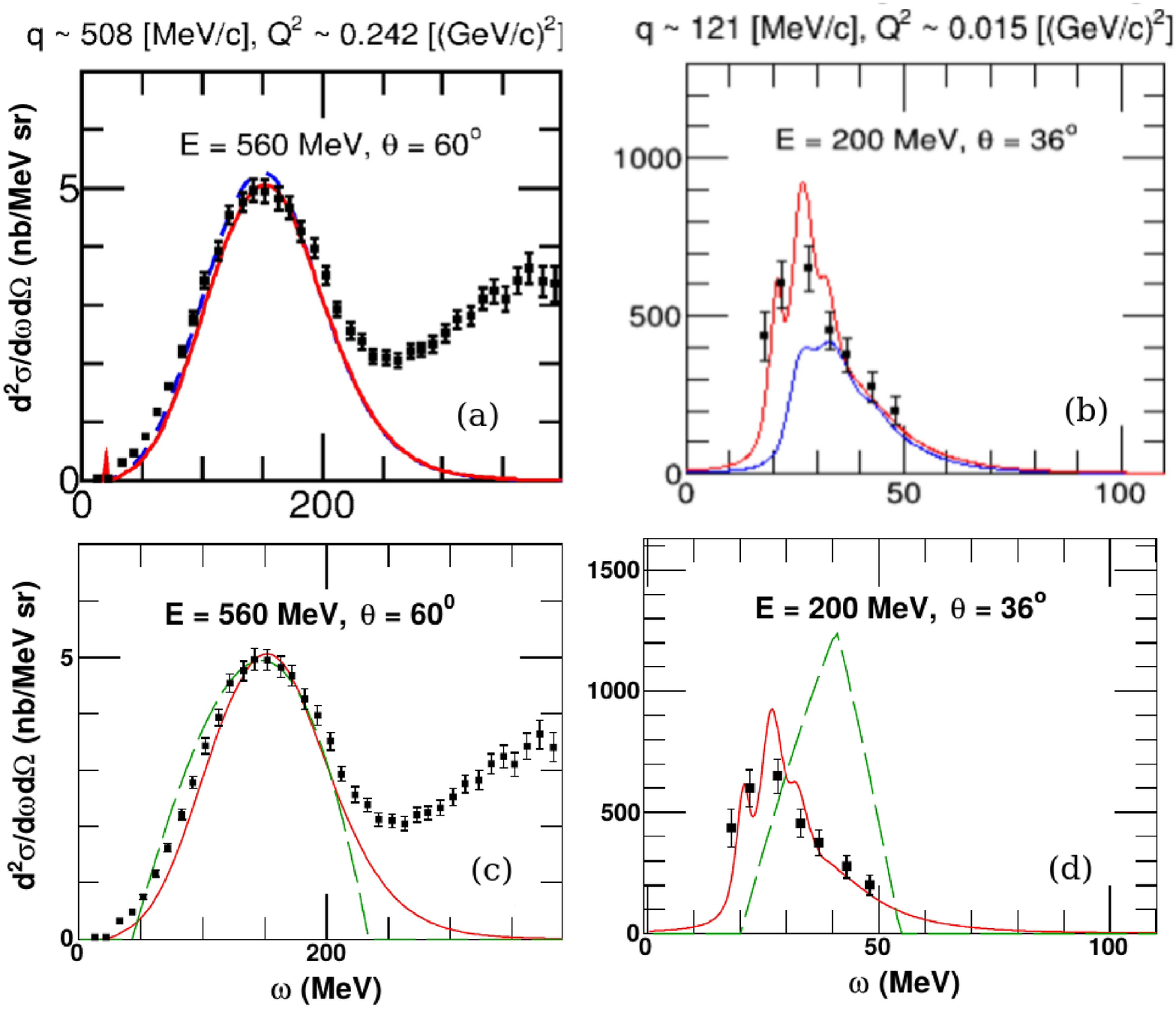}\vline
      \includegraphics[width=0.45\textwidth,angle=0]{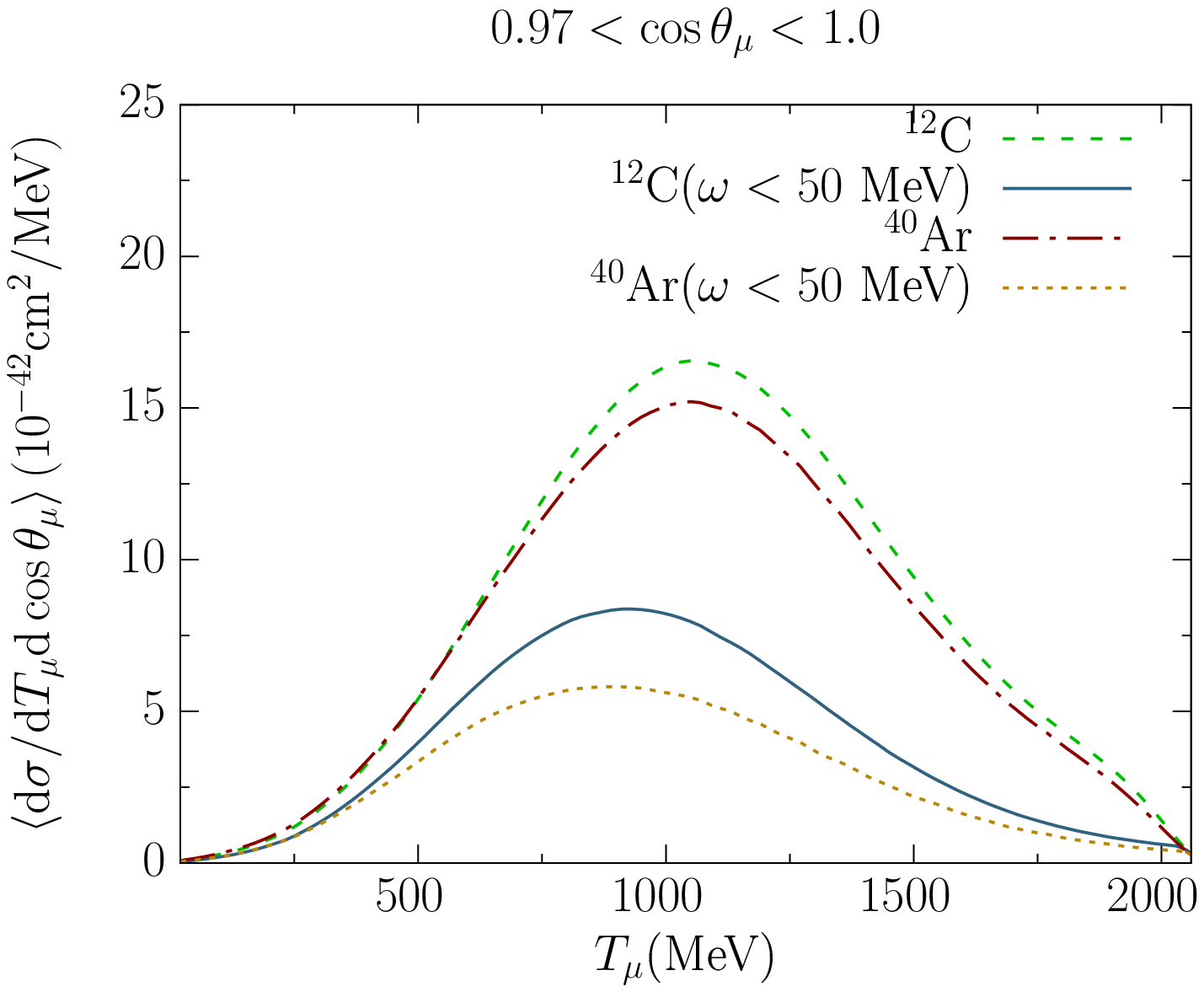}
  \caption{ Left panels: Inclusive $^{12}C(e,e')$ data~\cite{Sealock89,Barreau83} are compared with different model predictions. 
  In panels (a)-(b), we show HF and CRPA results. In panels (c)-(d), we compare RFG and CRPA predictions. Figures adapted from Refs.~\cite{Pandey15,Jachowicz16,Pandey-PhD}. Right panel: CRPA result for the MicroBooNE flux-folded double-differential cross section for CCQE neutrino-$^{40}Ar$ and $^{12}C$ scattering, at forward scattering angles. The low-energy contribution ($\omega<50$ MeV) is shown separately. Figure adapted from Ref.~\cite{VanDessel17}.}
  \label{fig:HF-CRPA-RFG}
\end{figure}

The model has been benchmarked against electron scattering data.
In panels (a) and (b) of Fig.~\ref{fig:HF-CRPA-RFG}, we show the effect of long-range correlations by comparing the Hartree-Fock (HF) with CRPA results. The predictions are contrasted with $^{12}C(e,e')$ data. The effect of long-range correlations is important at low-excitation energies [panel (b)], notably improving the agreement with data. Contrary, they induce only small corrections to the `bare' mean-field result at pure QE kinematics [panel (a)]. 
In the bottom panels, the CRPA predictions are compared with the relativistic global Fermi gas (RFG) model~\cite{Amaro05b,Gonzalez-Jimenez14b}.
Distortion effects in both initial and final nucleon wave functions, which are included in the HF and CRPA approaches but not in the RFG model, are important at small $Q^2$ [panel (b) and (d)]. Also, they are responsible for the tails observed above and below the QE peak [panel (c)]. This comparison is interesting because Fermi-gas based models are employed in many of the Monte Carlo neutrino event generators that are used to extract the neutrino oscillation probability from neutrino data. 

It is, therefore, clear that a proper description of the low-energy contributions ($\omega<50$ MeV) needs sophisticated nuclear modeling. At forward scattering angles, these low-energy contributions contribute to a good amount of the total strength. This is shown in the right panel of Fig.~\ref{fig:HF-CRPA-RFG}, where we present the CRPA predictions for the single-differential CCQE neutrino-$^{40}Ar$ cross section, folded with the MicroBooNE flux. For forward angles, the strength from $\omega<50$ MeV is approximately $30-50\%$ of the total. Similar results are found for the T2K flux~\cite{Pandey16} at similar kinematics. 

\subsection{Two-nucleon knockout mechanisms: SRC and MEC }

In our approach, two-nucleon knockout processes are induced by means of short-range correlations (SRC) and meson-exchange currents (MEC). Here, we comment on some relevant aspects of our model, further details can be found in~\cite{VanCuyck16,VanCuyck17}. 

The electroweak current operator $\hat{J}$ contains the one-body and the MEC operators, $\hat{J}=\hat{J}_{1}+\hat{J}_{mec}$. 
Short range correlations are introduced by applying a correlation operator $\hat{G}$, which contains central, spin-isospin and tensor parts, to the uncorrelated nuclear wave function $|\Phi\rangle$: 
$|\Psi\rangle\sim\hat{G}|\Phi\rangle$, with $|\Psi\rangle$ the correlated wave function.
The complexity introduced by the SRCs is then shifted to the current operator, 
which results in an effective current operator $\hat{J}_{eff}\sim\hat{G}^\dagger\left(\hat{J}_{1}+\hat{J}_{mec}\right)\hat{G}$. 
This allows us to consistently account for the SRC-MEC interference terms. 
We also stress that in our approach i) initial and final nucleons are HF mean-field wave functions, i.e., they are bound and scattering solutions of the Schr\"odinger equation in the same mean-field potential; and 
ii) we calculate the effect of SRC and MEC in both the one-nucleon knockout and two-nucleon knockout responses. 

\begin{figure}[htbp]
  \centering  
      \includegraphics[width=0.65\textwidth,angle=0]{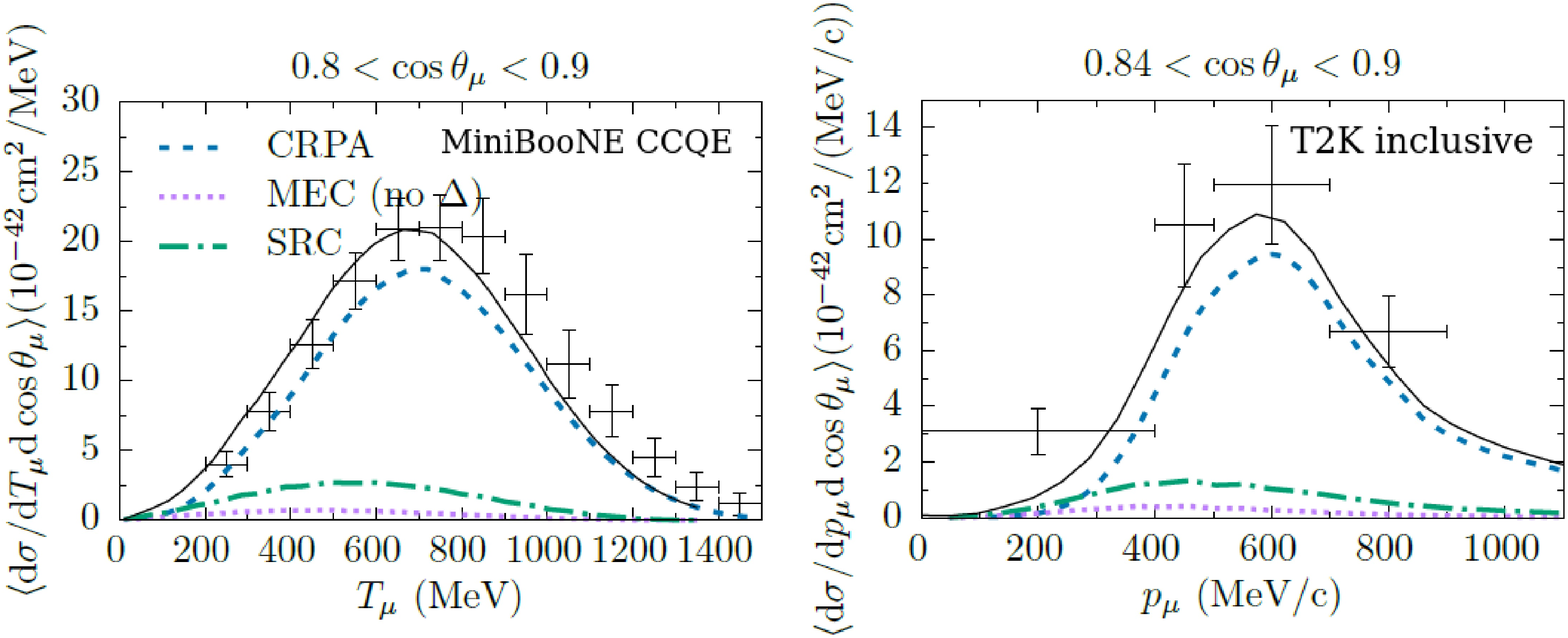}
  \caption{ MiniBooNE CCQE (left) and TK2 inclusive (right) double differential cross sections are compared with our predicitons. Data from~\cite{MiniBooNECC10,T2Kinc13}. Figures adapted from Ref.~\cite{VanCuyck17}.}
  \label{fig:MB-T2K}
\end{figure}

In Fig.~\ref{fig:MB-T2K} we present the double differential cross sections folded with the MiniBooNE and T2K fluxes. 
We have shown separately the one-nucleon knockout response (CRPA), and the two nucleon knockout responses (MEC and SRC). 
Delta currents are not yet included in the MEC contributions. 

\subsection{Single-pion production}

Single-pion production cross sections are described within the Hybrid-RPWIA model presented in Refs.~\cite{Gonzalez-Jimenez17a,Gonzalez-Jimenez17b}.
The starting point is the description of the elementary reaction with a microscopic low-energy model similar to that of Ref.~\cite{Hernandez07}, which includes resonances and background contributions. This low-energy model is combined with a Regge approach that provides the right behavior of the scattering amplitude at high energies. The current operator of the elementary reaction is then included in a nuclear framework by using Relativistic Mean-Field (RMF) wave functions for the bound nucleons~\cite{Gonzalez-Jimenez17b,Praet09}.

\begin{figure}[htbp]
  \centering  
      \includegraphics[width=0.3\textwidth,angle=0]{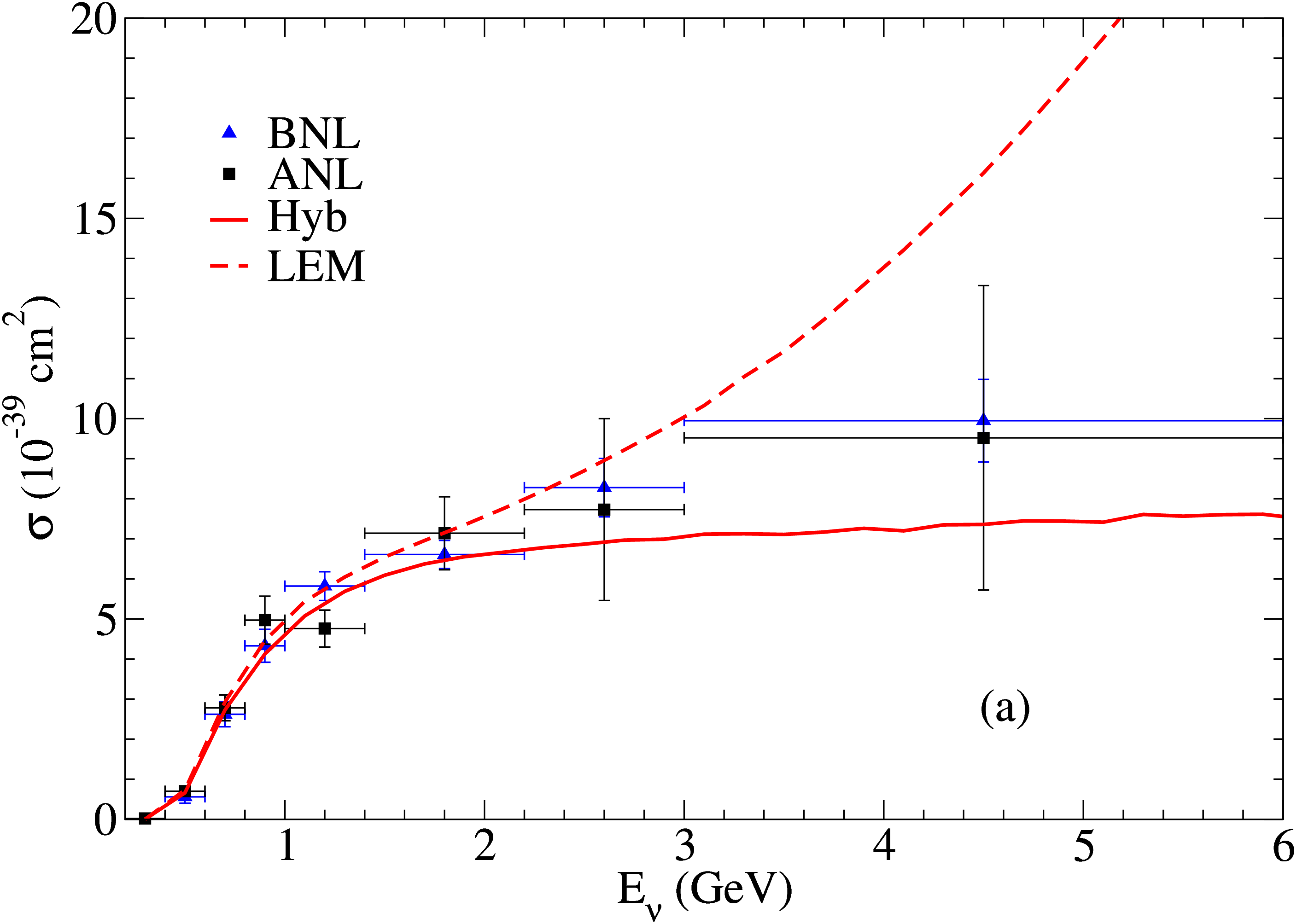}
      \includegraphics[width=0.3\textwidth,angle=0]{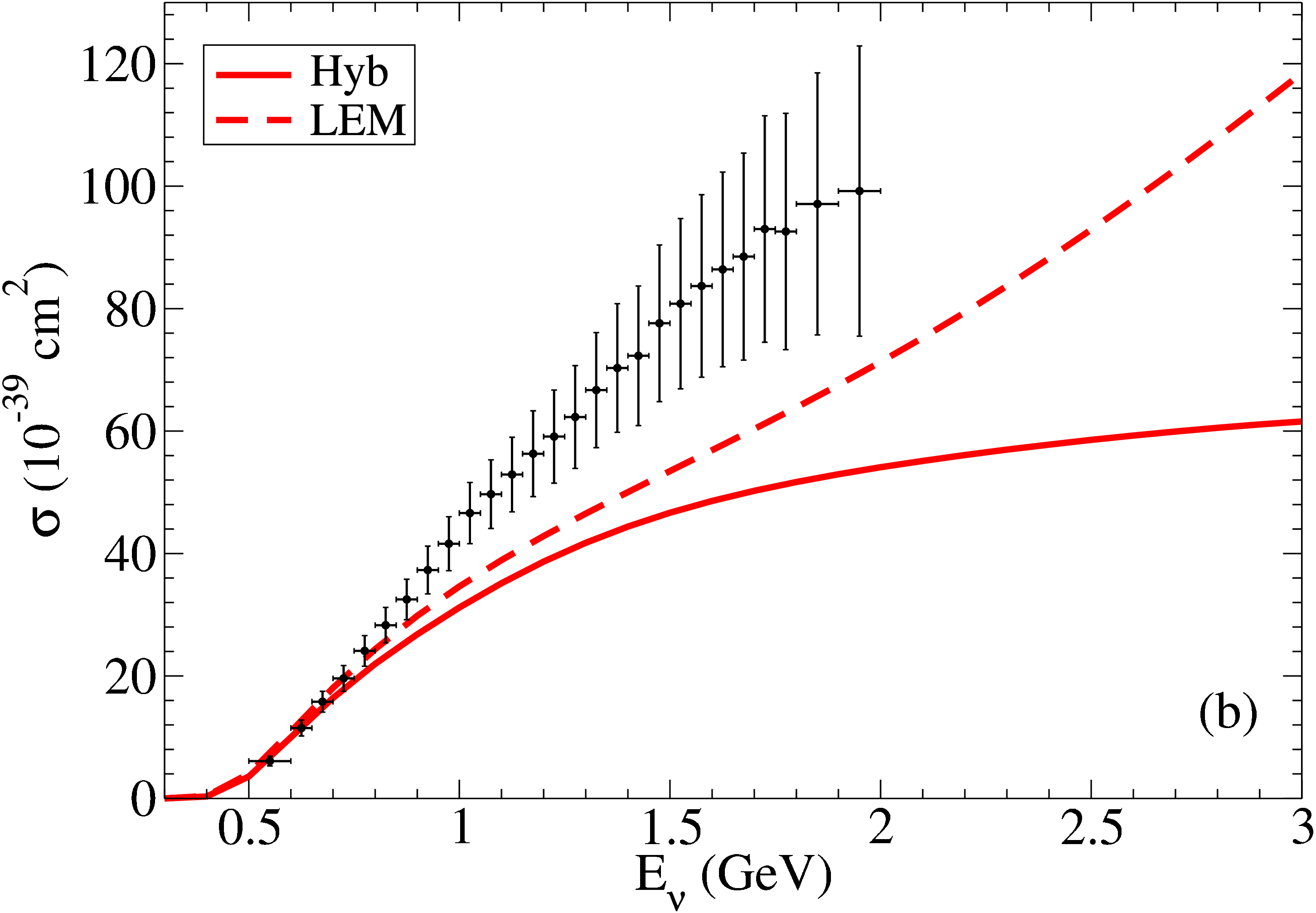}\\
      \includegraphics[width=0.3\textwidth,angle=0]{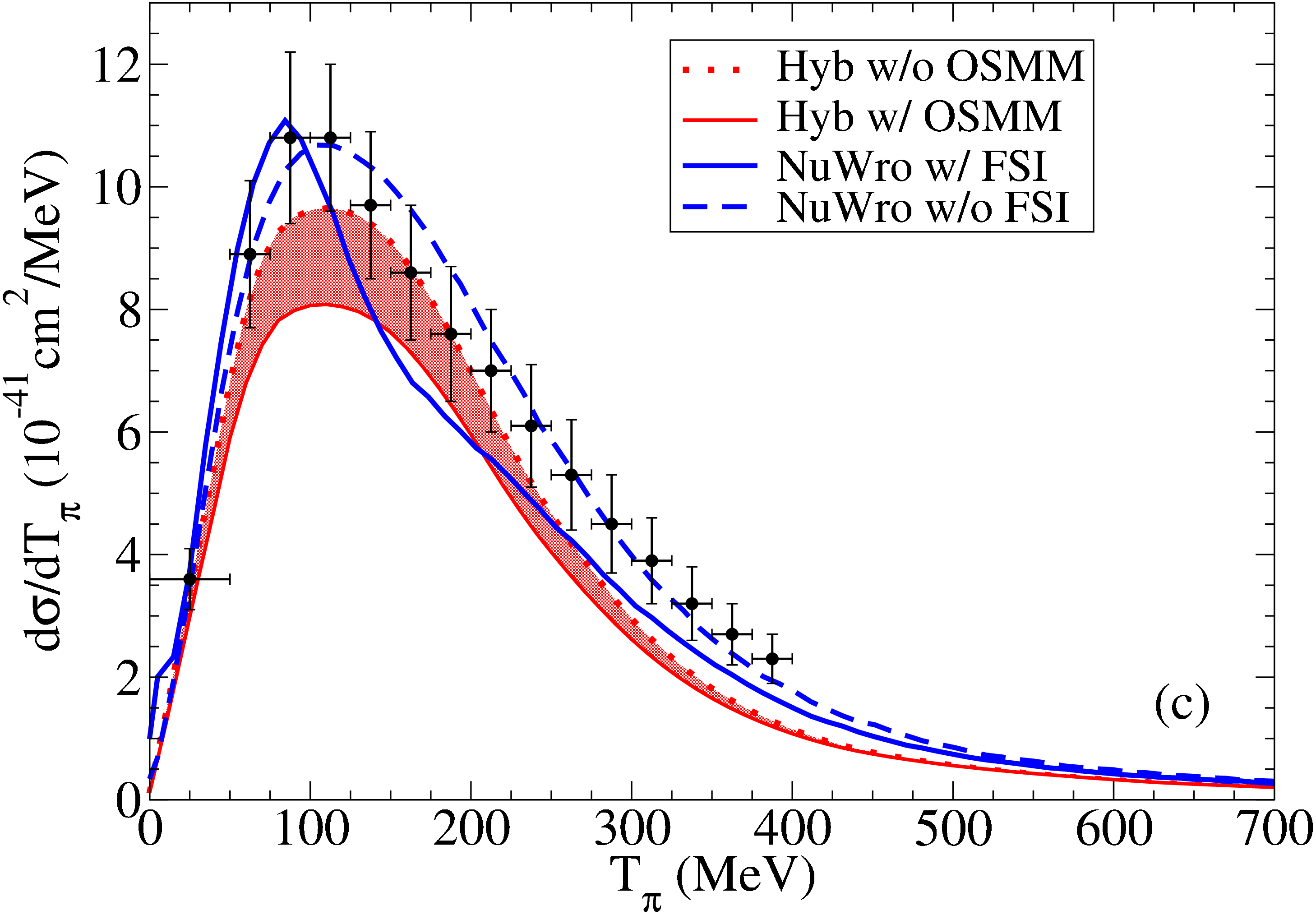}
      \includegraphics[width=0.3\textwidth,angle=0]{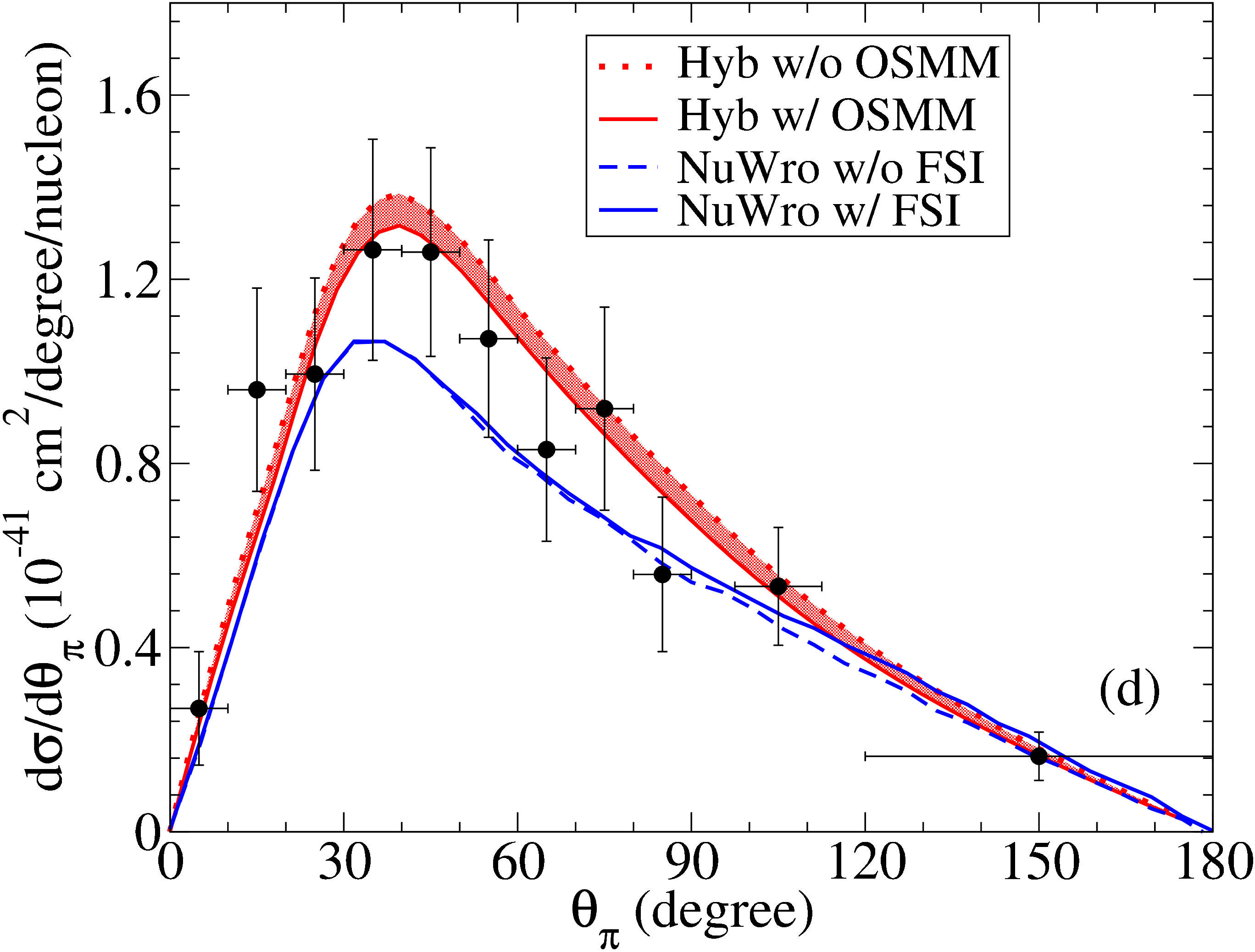}      
  \caption{ Panels (a) and (b) show the total cross section computed with the Hybrid and low-energy models for the CC $\nu$-induced 1$\pi^+$ production on proton [panel (a), data from~\cite{Wilkinson14}] and on the MiniBooNE $CH_2$ target [panel (b), data from~\cite{MINERvACCpi16}]. Panel (c) is the flux-folded single differential cross sections for the MiniBooNE $\nu$CC 1$\pi^+$~\cite{MBCCpionC11} and MINERvA $\bar\nu$CC 1$\pi^0$~\cite{MINERvACCpi16} samples.  
  Figures adapted from Refs.~\cite{Gonzalez-Jimenez17a,Gonzalez-Jimenez17b}.}
  \label{fig:pions}
\end{figure}

We summarize its main features as follows. 
(i) The process is described in a fully relativistic framework. The nucleon bound-state wave are RMF wave functions, therefore, in-medium effects like Fermi motion and nuclear binding are consistently included. 
(ii) Since the formalism works at the amplitude level, it can provide predictions for complete kinematics. 
(iii) The high-energy behavior is dictated by Regge phenomenology, this way curing the pathological behavior often observed when low-energy models are extended to a higher energy regime. This is illustrated in panels (a) and (b) of Fig.~\ref{fig:pions} by comparing the results from the low-energy model (dashed lines) with the ones from the Hybrid model (solid lines). 
Panel (a) shows our prediction for the $\nu$-induced 1$\pi^+$ production on a hydrogen target, while panel (b) presents the results for the same reaction channel but with the MiniBooNE target CH$_2$.

FSI are not taken into account in the Hybrid-RPWIA model, work is in progress to amend this. To judge the effect of FSI on the cross sections we study the results from the NuWro Monte Carlo (MC) event generator~\cite{NuWro-web} calculated with and without FSI. 
This is shown in panels (c) and (d) of Fig.~\ref{fig:pions} for the single-differential cross section folded with MiniBooNE and MINERvA fluxes, respectively. The lower limit of the red band corresponds to the calculation when the delta-decay width is modified to account for in-medium effects, according to the Oset and Salcedo prescription~\cite{Oset87}. The upper limit is the calculation with the free decay width.

\vspace{1cm}
Summarizing, we have presented an overview of the recent developments of the Ghent group on the different reaction mechanisms involved in neutrino-nucleus interaction at intermediate energies. 
Work is in progress to complete the MEC calculation by including the delta currents, and to implement FSI in the pion-production model.\\

This work was supported by the Interuniversity Attraction Poles Programme initiated by the Belgian Science Policy Office (BriX network P7/12), the Research Foundation Flanders (FWO-Flanders), and the Special Research Fund, Ghent University. The computational resources (Stevin Supercomputer Infrastructure) and services used in this work were provided by Ghent University, the Hercules Foundation and the Flemish Government.
J.N. was supported as an `FWO-aspirant'.
V.P. acknowledges the support by the National Science Foundation under Grant No. PHY-1352106.
K.N. was partially supported by the Polish National Science Center (NCN), under Opus Grant No. 2016/21/B/ST2/01092, as well as by the Institute of Theoretical Physics, University of Wroc{\l}aw Grant No. 0420/2545/17.

\linespread{1.}

\bibliographystyle{apsrev4-1}
\bibliography{bibliography}

\end{document}